# Boosted-SpringDTW for Comprehensive Feature Extraction of Physiological Signals

Jonathan Martinez, *Student Member, IEEE*, Kaan Sel, *Student Member, IEEE,* Bobak J. Mortazavi, *Member, IEEE*, and Roozbeh Jafari, *Senior Member, IEEE*

*Abstract— Goal*: To achieve high-quality comprehensive feature extraction from physiological signals that enables precise physiological parameter estimation despite evolving waveform morphologies. *Methods*: We propose Boosted-SpringDTW, a probabilistic framework that leverages dynamic time warping (DTW) and minimal domain-specific heuristics to simultaneously segment physiological signals and identify fiducial points that represent cardiac events. An automated dynamic template adapts to evolving waveform morphologies. We validate Boosted-SpringDTW performance with a benchmark PPG dataset whose morphologies include subject- and respiratory-induced variation. *Results*: Boosted-SpringDTW achieves precision, recall, and F1-scores over 0.96 for identifying fiducial points and mean absolute error values less than 11.41 milliseconds when estimating IBI. *Conclusion*: Boosted-SpringDTW improves F1-Scores compared to two baseline feature extraction algorithms by 35% on average for fiducial point identification and mean percent difference by 16% on average for IBI estimation. *Significance*: Precise hemodynamic parameter estimation with wearable devices enables continuous health monitoring throughout a patients' daily life.

*Impact Statement—* Boosted-SpringDTW enables the simultaneous segmentation of physiological signals into cardiac cycles and identification of all desired fiducial points that will lead to high-quality health parameter estimation.

## I. Introduction

HEMODYNAMIC parameter estimation with wearable devices enables continuous health monitoring in outpatient settings, thus granting diagnostic insight throughout varying contexts of a patient's daily activities. Not commonly captured in clinical settings, this information contributes to the early detection of life-threatening illnesses, advanced fitness tracking, and emotional regulation [1]. Precise parameter estimation depends on comprehensive feature extraction from a stream of physiological signals continuously captured by conveniently worn wearable devices. Furthermore, features are extracted from distinguishable peaks, valleys, and slopes within the waveform that depict key physiological events. For example, photoplethysmography (PPG) waveform morphology captures blood flow, which may be mapped to heart rate (HR), interbeat interval (IBI), blood pressure (BP), and respiration rate (RR) [2]. However, physiological waveform morphology is highly sensitive to inter-subject and contextual variations that present challenges to the heuristics pre-defined by previously developed feature extraction frameworks [3]. This variability may lead to inaccurate parameter estimation, ultimately leading to unreliable health diagnoses. In this study, we introduce a comprehensive feature extraction framework that adapts to such morphological variations, ensuring high-quality hemodynamic parameter estimations.

All physiological waveforms that capture blood flow or represent heart beats – including photoplethysmography (PPG) [4], electrocardiography (ECG) [5], and bio-impedance (Bio-Z) [6] – possess a quasi-periodic property corresponding to the contraction and relaxation of the heart. Typical approaches for parameter estimation and analysis, beginning with segmentation, often revolve around hand-crafted fiducial point detection algorithms that are constrained by domain-specific heuristics such as average cycle length or plausible amplitude values [7]–[9]. Unless adaptive thresholding is carefully implemented to consider all possible sources for morphological variation, over time as patients age, vascular health evolves, or as the algorithms are applied to new patients, such methods will fail when faced with a variation that is not considered in the general case. Filtering and transformation of the signals to the frequency domain, such as with Hilbert or wavelet transforms [10]–[12], are more robust, however, they still depend on effective adaptive thresholding for task-specific identification algorithms. Alternatively, machine learning and deep learning [13]–[15] models do not require adaptive thresholding and are robust to waveform variations, but require an abundant labelled training dataset with high waveform variance that is not practical to obtain for patients whose data was not seen in

J. Martinez is with the Department of Computer Science and Engineering, Texas A&M University, College Station, TX, 77840 USA (e-mail: jmartinez0304@tamu.edu).
K. Sel is with the Department of Electrical and Computer Engineering, Texas A&M University, College Station, TX, 77840 USA (e-mail: ksel@tamu.edu)
B. Mortazavi is with the Department of Computer Science and Engineering, Texas A&M University, College Station, TX, 77840 USA (e-mail: bobakm@tamu.edu)
R. Jafari is with the Department of Biomedical Engineering, Electrical and Computer Engineering, Computer Science and Engineering, Texas A&M University, College Station, TX, 77840 USA (e-mail: rjafari@tamu.edu)

training.

Dynamic time warping (DTW) shows potential for simultaneously identifying all target fiducial points. This established technique compares the likeness between two signals to produce sample-to-sample mappings between them without any prior knowledge on the their underlying physics [16]. Thus, it has been used for subsequence matching, feature extraction, clustering, optimization, and signal quality index (SQI) tasks [17]–[19]. Although it is a good candidate for the objective of this work, in its original form DTW faces two critical challenges when segmenting a stream of data. First, quasi-periodic physiological waveforms contain multiple meta-subsequences that resemble the morphology of a complete cardiac cycle [20], which is detrimental to segmentation tasks since incomplete cycles might be detected. Second, when DTW depends solely on a single template, waveform comparison quality will decline when encountering extreme morphological variations. Although additional innovations have been proposed to encode signal characteristics to a feature vector that may overcome morphological variations [21], these approaches become impractical to determine the optimal feature sets to be used for DTW comparisons.

In this work, we propose Boosted-SpringDTW to overcome the above challenges when performing comprehensive feature extraction sub-tasks for a stream of quasi-periodic physiological signals: 1) segmentation into cardiac cycles and 2) identification of all fiducial points. Given a single template of a typical cardiac cycle, our framework conducts a probabilistic decision-making process to precisely identify the true start and endpoints within a waveform stream by leveraging both minimum domain-specific characteristics of the physiological signal and the generalizable intuitions of DTW. We also propose a dynamic template that will automatically adapt to new variations without domain expert intervention.

Our contributions in this paper are summarized as follows:
- We introduce Boosted-SpringDTW which simultaneously identifies all fiducial points of a given waveform stream.
- We propose a probabilistic decision-making process that enhances DTW with minimal domain-specific heuristics, thus enabling the analysis of quasi-periodic signals.
- We incorporate an automated dynamic template to adapt to evolving morphologies.

II. MATERIALS AND METHODS

Boosted-SpringDTW – shown in Fig. 1 – achieves comprehensive feature extraction through two sub-tasks solved simultaneously: 1) cardiac cycle detection (segmentation) and 2) fiducial point identification. The segmentation task leverages minimal prior domain-knowledge of the target signal to first identify all realistic candidate endpoints for each cardiac cycle. Then, the true endpoints are distinguished by tracking DTW distance scores within a reasonable search space constrained by HR. Fiducial points are simultaneously identified based on the sample-to-sample mappings. New templates are automatically generated over time as the waveform morphology evolves. In this study, we apply Boosted-SpringDTW to streams of PPG waveforms, however, this approach is applicable to all types of physiological waveforms.

*A. Cardiac Cycle Detection & Fiducial Point Identification*

Segmenting physiological signals into cardiac cycles involves assigning a probability, $P(e_t)$ to each sample, $x_t$, in the given stream of PPG waveform, $X$. This represents the likelihood in which each sample is either a start or endpoint, $e_t$, of a cardiac cycle (segment) computed as

$$P(e_t) = P(c_t|x_t) * P(d(x_t, y_m)) \quad (1)$$

where $P(c_t)$ is the likelihood based on the minimal domain-specific heuristics which consider the underlying physics of the waveform where $c_t$ represents a candidate endpoint in $X$, and $P(d(x_t, y_m))$ is the likelihood based on morphology similarity

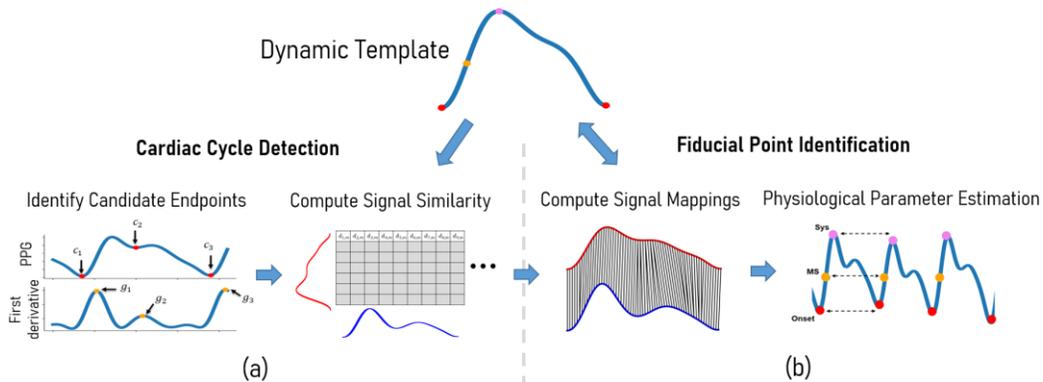

Fig. 1. Overview of the proposed Boosted-SpringDTW framework that achieves comprehensive feature extraction through cardiac cycle segmentation and fiducial point identification, a) first we combine domain-specific heuristics with morphology-based comparisons from DTW to achieve cardiac cycle segmentation where $c_t$ are candidate endpoints, $g_t$ are the max slope amplitudes used to rank candidate endpoints, and $d_{t,m}$ is the DTW distance between the stream and template waveforms, and b) we use the DTW sample-to-sample mappings to identify fiducial points – Onset is the cardiac cycle endpoint, MS is the max slope point, and Sys is the systolic peak. We also introduce of a dynamic template that may adapt to evolving morphologies

evaluated by DTW where $d(x_t, y_m)$ is the DTW distance between a given sample of X, $x_t$, and a given sample of the template waveform $Y$, $y_m$. Furthermore, the DTW distances that are typically monitored for subsequence matching in the standard SpringDTW approach are "boosted" with the domain-specific heuristic likelihood score to bring forward a probabilistic decision-making process. We describe the computation of each in the following paragraphs.

We detect the local maxima of a series of likelihood scores to identify the time steps that are the true cardiac cycle endpoints. Local maxima are defined by the dominant frequency, $f^*$, of the pulsatile signals, which corresponds to HR. Using a standard Fast Fourier Transform (FFT) [22] to obtain , $f^*$, we estimate the average cardiac cycle length, $l_X$, for windows of the input batch with:

$$l_X = \frac{f_s}{f^*} \quad (4)$$

where $f_s$ is the sampling rate of $X$. In this study we used 1-minute batches to compute the average cycle length. The batch length impacts the precision of the estimated $l_x$, where a smaller length yields precise estimations robust to fast changes in HR. However, this requires frequent executions of FFT thus increasing the overall runtime of the framework. On the other hand, there should be at least two complete cardiac cycles present in the batch to prevent detecting dominant frequencies of incomplete meta-subsequences.

$P(c_t|x_t)$ is the likelihood that the sample, $x_t$, is a realistic, candidate endpoint for the type of waveform being analyzed based on understanding of the physiological processes that compose it. For our experiments we use PPG, where we understand that start and endpoints for all cardiac cycles can be characterized by the onset point that immediately precedes the systolic fiducial point. Therefore, the set of candidate endpoints, $C$, within a given batch $X$ includes all local minima and will be scored based on the steepness of the immediately following peak. (Discussion of cardiac behavior represented by PPG included in Supplementary I.A.) This is formulated accordingly:

$$P(c_t|x_t) = \frac{g_t - \min(X\prime)}{\max(X\prime) - \min(X\prime)} \quad (2)$$

where $g_t$ is the value of the max slope point represented in the first derivative as a peak point, as shown in Fig. 2. All values in the first derivative of PPG are scaled to where the maximum value is 1 and the minimum value is 0 where the greater the max slope of the onset following a candidate endpoint, the more likely that it is a true endpoint. Understanding of cardiac cycle endpoint features is the only prior domain-knowledge required for this framework, and this concept is generalizable to all types of physiological waveforms that measure blood flow.

$P(d(x_t, y_m))$ is the morphology-based likelihood – derived from DTW – that each sample $x_t$ in $X$ is a candidate endpoint based on the comparison of each sample in the input stream to all points in the template $y_m$ in $Y$, to compose a running DTW distance matrix, $D(X,Y)$. Background material on DTW included in Supplementary I.B. Since the Euclidean distance of each sample-to-sample comparison is augmented with the minimum distance most adjacent (or most recently computed) in the distance matrix, we can consider the distance function as causal, making the most recent comparison at any given $t$, $d(x_t, y_m)$, to be the most representative metric for similarity between $X$ and $Y$ at that time. Furthermore, the smaller this distance value becomes, the more likely the region we are analyzing is a subsequence (or cardiac cycle). The likelihood score is computed as follows:

$$P(d(x_t, y_m)) = e^{-\gamma * d(x_t, y_m)} \quad (3)$$

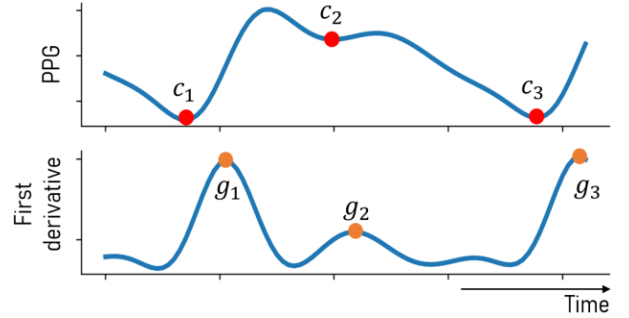

Fig. 2. Candidate endpoints are identified in the raw PPG signal and are assigned a likelihood score based on the gradient of the following peak.

We leverage the exponential function as it is a monotonically increasing function that will penalize large $d(x_t, y_m)$ distance values appropriately. We also use $\gamma$ as a scaling factor to normalize the distribution of likelihood scores to be easily comparable with $P(c_t|x_t)$, this scaling factor should be determined empirically while considering the scale of DTW distance values and the amount of influence which the resulting $P(d(x_t, y_m))$ should carry on the final prediction task.

The search for true cardiac cycle endpoints occurs as DTW distances are computed. The process begins at the first potential $c_t$, where we expect a corresponding $e_t$ to exist around the next $l_X$ time steps of the stream with the greatest $P(e_t)$. By defining two generalizable parameters $\alpha$ and $\beta$, we can define the amount of tolerance surrounding $l_X$ that we will allow to detect an optimal $e_t$. Here, $\alpha * l_X$ will be the minimum distance in time steps after a potential $c_t$, therefore if another candidate endpoint occurs between the first candidate and $l_\alpha = c_t + \alpha * l_X$, then the segmentation process will reset and this new point will become the new potential cardiac cycle starting point. Then, $\beta * l_X$ will be the maximum distance after $c_t$ which a valid $e_t$ may exist, $l_\beta = c_t + \beta * l_X$, where we will accept the candidate endpoint with the greatest $P(e_t)$ between $l_\alpha$ and $l_\beta$ as the true $e_t$. If no candidate endpoint is detected in this region, then the algorithm will reset to the next candidate endpoint after $l_\beta$. Here, $\alpha$ and $\beta$ may be tuned to impact sensitivity of the framework but ideally the minimum distance should be large enough to avoid the possibility of false positives (such as detecting a dicrotic notch of the current cardiac cycle as $e_t$ in PPG), and the maximum distance should be greater than $l_X$ yet less than the potential location of non-endpoint fiducial points (such as the dicrotic notch in PPG) in the subsequent cardiac cycle. This approach may be visualized in Fig. 3.

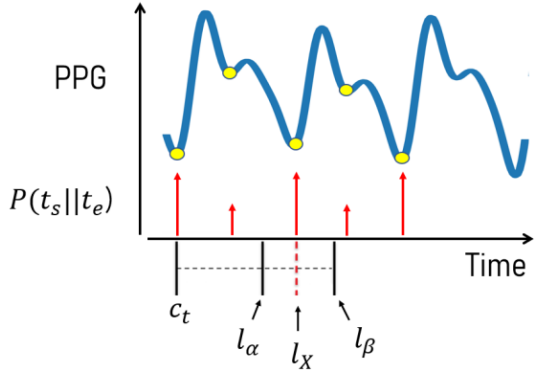

Fig. 3. True cardiac cycle endpoints are identified as local maxima in $P(t_s||t_e)$ values within the search space constrained by an average cycle length, $l_X$, which is derived from local HR.

*B. Automated Dynamic Template*

We define the maximum number of templates allowed in the ensemble, $k$, and initialize it with a prime template, $Y^*$. Then, analysis of an input waveform batch begins with the prime template only. After feature extraction concludes for a region of the batch with a predefined length of $u$, if there are less than $k$ templates in the ensemble then a new template will be generated and added to it. This is accomplished by generating a consensus sequence based on the cardiac cycles detected in this region - using DTW barycenter averaging [23] – and its fiducial points will be annotated by the prime template. Further explanation of the composition of the consensus sequence is provided in Supplementary I.C. While multiple templates exist in the ensemble, each will conduct analysis over the region independently and the average warping path distance, $|w_{X,Y}|$, between each detected cardiac cycle and each template will be tracked. When all templates have completed their analysis, the template which yielded the lowest average $|w_{X,Y}|$ will be predicted as the optimal template, $Y_{opt}$, which has achieved the most accurate feature extraction for the set of detected cardiac cycles. Last, for the case that there are $k$ templates in the ensemble and $Y^* \neq Y_{opt}$, then an update will be triggered and the least frequently used template will be discarded from the ensemble. Otherwise, if $Y^* = Y_{opt}$, then an update will not be triggered. This template update protocol is possible due to the robustness of our cardiac cycle detection phase to at the least be capable of identifying the true start and endpoints despite that the fiducial point mapping could potentially be low-quality, where an additional pass over the most recent segment may be conducted with the newly generated template to achieve fiducial point identification with greater precision at the small cost of the additional time required to re-analyze the input batch.

## III. RESULTS

We validate the effectiveness of Boosted-SpringDTW (single and dynamic template) to achieve comprehensive feature extraction from PPG waveforms that are impacted by subject- and respiration-induced morphological variations in the IEEE TBME Respiratory Rate Benchmark Data Set [24]. The dataset includes 42 participants –29 children and 14 adults. Signals are sampled at 300 Hz, and the PPG signals were preprocessed with a 4[th] order Butterworth Bandpass Filter with cutoff frequencies of [0.5, 5]. We empirically determined Boosted-SpringDTW parameters $\alpha$, $\beta$, and $\gamma$ to be set to 0.7, 1.3, and 5000. We compare performance to two baselines – the original SpringDTW algorithm [25] and adaptive thresholding. SpringDTW detects subsequence endpoints by tracking minimum $d(x_t, y_m)$ distances over a stream. Adaptive thresholding leverages amplitude and distance-based heuristics to detect peak, slope, and valley features with PPG and its derivative signals. [26], [27]. We then evaluate IBI estimation with each of the identified fiducial points. To maintain a consistent problem setting in this study, we analyze only solutions that operate in the time-domain and that do not require training data to tune model parameters.

*A. Fiducial Point Identification*

Fig. 4 shows the fiducial points identified for this experiment that represent key events in the cardiac cycle [4]. We formulate both a classification and a regression task where each sample in a PPG stream is labelled as either a systolic peak (SYS), max slope point (MS), cardiac cycle endpoint (EP), or a non-fiducial point (NF); and precision is based on the reported timestamp. All fiducial points were manually labeled with the assistance of a python-based graphical user interface [28], yielding 27850 SYS points, 27843 MS points, 27926 EP points, 5964423 NF points. The evaluation metrics include precision, recall, F1-score, and root mean squared error (RMSE) which are computed as follows

$$Precision = \frac{TP}{TP + FP} \qquad (9)$$

$$Recall = \frac{TP}{TP + FN} \qquad (10)$$

$$F1 = 2 \cdot \frac{Precision \cdot Recall}{Precision + Recall} \qquad (11)$$

$$RMSE = \sqrt{\frac{1}{F}\sum_{i=1}^{F}(\widehat{T_i} - T_i)^2} \qquad (12)$$

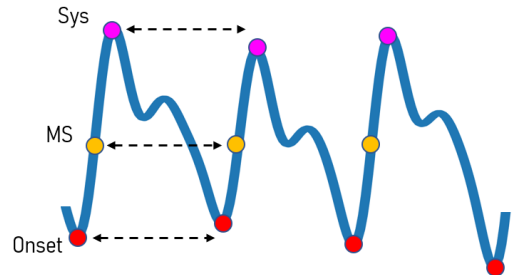

Fig. 4. Fiducial points to be identified and used for IBI estimation. The onset points represent the start and end for each cardiac cycle, the max slope point gives insight into the rate of change in blood pressure as the heart contracts, and the systolic peak is directly correlated to the maximum pressure of the blood flow through this region of the body after it is pumped from the heart. Each of these fiducial points are commonly used in the estimation of several physiological parameters, including IBI, HR, HRV, BP, RR, and others.

TABLE I
FIDUCIAL POINT IDENTIFICATION PERFORMANCE

|  | Sys | | | | MS | | | | EP | | | |
|---|---|---|---|---|---|---|---|---|---|---|---|---|
|  | Prec | Rec | F1-Score | RMSE (ms) | Prec | Rec | F1-Score | RMSE (ms) | Prec | Rec | F1-Score | RMSE (ms) |
| SpringDTW | 0.939 | 0.433 | 0.592 | 25.1 | 0.998 | 0.460 | 0.630 | 24.2 | 0.0 | 0.0 | 0.0 | 186.1 |
| Adaptive Thresholding | 0.996 | 0.789 | 0.880 | 42.6 | 0.994 | 0.790 | 0.881 | 44.3 | 0.688 | 0.995 | 0.813 | 166.9 |
| Boosted-SpringDTW-ST | 0.981 | 0.967 | 0.974 | 57.2 | 0.963 | 0.949 | 0.956 | 55.0 | 0.986 | 0.986 | 0.986 | 34.5 |
| Boosted-SpringDTW-DT | 0.995 | 0.980 | 0.987 | 34.9 | 0.979 | 0.965 | 0.972 | 36.2 | 0.986 | 0.986 | 0.986 | 34.5 |

TABLE II
IBI ESTIMATION PERFORMANCE

|  | Sys | | MS | | EP | | Valid Predictions, # (%) |
|---|---|---|---|---|---|---|---|
|  | MAE, ms (%) | Corr | MAE, ms (%) | Corr | MAE, ms (%) | Corr |  |
| SpringDTW | 29.14 ± 1.93 (5.08) | 0.823 | 31.84 ± 2.02 (5.53) | 0.81 | 219.28 ± 9.93 (31.69) | -0.377 | 25,874 (66.8) |
| Adaptive Thresholding | 131.61 ± 3.24 (23.91) | 0.356 | 142.18 ± 3.27 (25.49) | 0.335 | 110.64 ± 2.25 (13.44) | 0.443 | 65,629 (87.1) |
| Boosted-SpringDTW-ST | 10.81 ± 0.44 (1.44) | 0.976 | 11.71 ± 0.41 (1.56) | 0.979 | 11.41 ± 0.27 (1.59) | 0.989 | 79,451 (99.9) |
| Boosted-SpringDTW-DT | 7.59 ± 0.14 (1.09) | 0.997 | 8.54 ± 0.15 (1.24) | 0.996 | 11.41 ± 0.27 (1.59) | 0.989 | 79,464 (99.9) |

where TP is a correctly labelled fiducial point, TN is a correctly labelled non-fiducial point, FP is an incorrectly labelled non-fiducial point, FN is an incorrectly labelled fiducial point, $F$ is the number of fiducial point observations, $T_i$ refers to the ground truth timestamp of a fiducial point, and $\hat{T}_i$ refers to the predicted timestamp of a fiducial point. These metrics were computed for each class of fiducial points independently. RMSE only compares time distances for positive predictions. Therefore, the number of samples may vary for each algorithm. Table I shows the scores for fiducial point identification performance for each algorithm included in the study where Boosted-SpringDTW-ST refers to using a single template and Boosted-SpringDTW-DT refers to using the dynamic template.

### B. Physiological Parameter Estimation

We evaluate the estimation of a well-studied physiological parameter from the detected fiducial points, IBI, which is highly regarded for health monitoring [29]. IBI is defined as the time difference between consecutive heart beats and is computed as $IBI_t = Sys_t - Sys_{t-1} = MS_t - MS_{t-1} = t_e - t_s$. Ground truth IBI was extracted from the dataset's ECG waveform R peaks – also manually annotated. We evaluate the estimation performance using mean absolute error (MAE) in milliseconds (ms) and also using Pearson's correlation between the closest predicted IBI value in time and the ground truth measurements with a maximum difference in reported time of 1 second. We also included a plausibility filter for the estimated IBI values where estimates that implied HRs below 40 beats per minute (IBI of 300 ms) or above 100 beats per minute (IBI of 1500 ms) were discarded. This yields 25,874, 65,629, 79,451, and 79,464 valid IBI estimates for SpringDTW, adaptive thresholding, and the two proposed Boosted-SpringDTW frameworks. Results shown in Table II indicate that fiducial point identification performance is directly linked to the quality of the resulting estimates for physiological parameters.

### IV. DISCUSSION

Both versions of the proposed Boosted-SpringDTW framework show strong performance for identifying each class of fiducial points achieving scores over 0.949 for precision, recall, and F1. The improved F1 compared to the two baseline algorithms are the best reflection of overall performance since they represent the balance in TP, FP, and FN predictions. The dynamic template also yields an additional 1-2% improvement labelling SYS and MS points. Although the lowest RMSE values were by SpringDTW, it yielded fewer positive predictions for each fiducial point.

In Fig. 5 we show dynamic templates generated for six subjects where the single template framework previously achieved F1-scores less than 0.95 for identifying MS points yet we observe scores over 0.98 with the dynamic template. This is due to the dynamic template update protocol that adapts to new variations in waveform morphology.

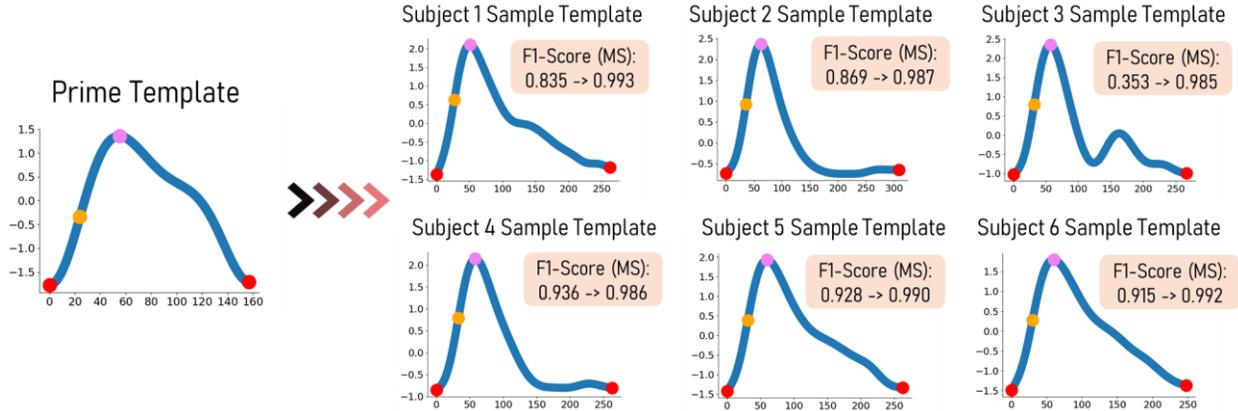

Fig. 5. Generated dynamic templates for six subjects where the single template case achieved F1-scores below 0.95 for identifying MS yet the dynamic template is able to achieve scores greater than 0.98. The dynamic template update protocol precisely captures intricacies of evolving waveform morphology.

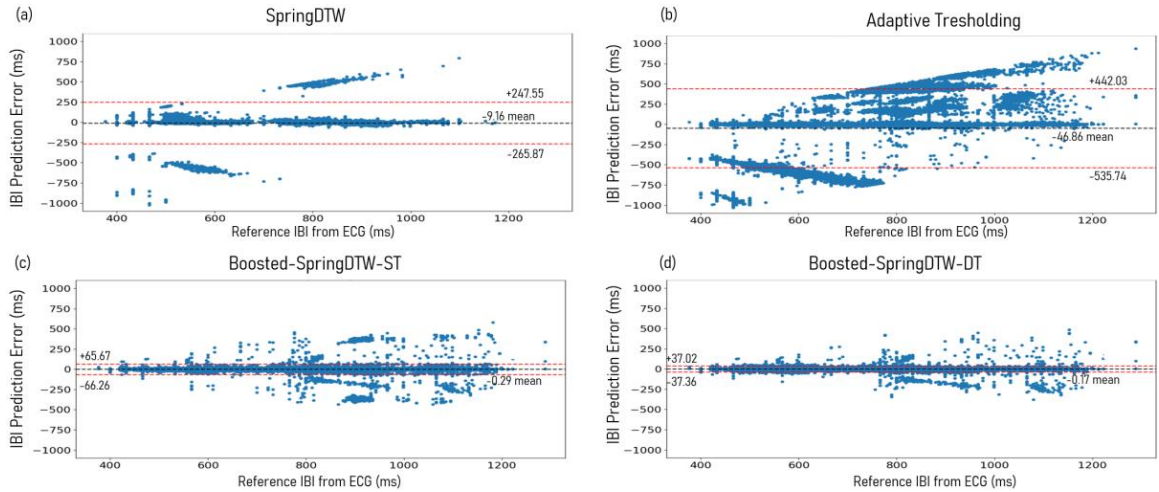

Fig. 6. Difference plots evaluating IBI predictions computing using all fiducial points detected by (a) SpringDTW, (b) adaptive thresholding, (c) Boosted-SpringDTW-ST, and (d) Boosted-SpringDTW-DT.

The top performers for IBI estimation with each fiducial point are the two versions of Boosted-SpringDTW yielding MAE scores less than 12 ms and correlation coefficients greater than 0.98. We observe improved IBI estimation with SYS and MS points when Boosted-SpringDTW leverages the dynamic template, yielding a 28.5% improvement in MAE and a 2.0% improvement in the correlation coefficient values. In the difference plots shown in Fig. 6-c and Fig. 6-d we observe a smaller margin of error for Boosted-SpringDTW compared to the two baseline approaches – observed in Fig. 6-a and 6-b – and noticeably less outlier predictions when using the dynamic template. Discussion on runtime analysis is in Supplementary I.D.

Fig. 7 shows the trend of $d(x_t, y_m)$ as it is computed over a PPG stream for each time step and the associated endpoint likelihood scores. The $d(x_t, y_m)$ distances are greatest at the very beginning of a waveform but will reach the local minima when the first possible subsequence is encountered. However, despite the existence of more optimal subsequences, the $d(x_t, y_m)$ values will saturate and gradually continue to increase since the distance at each subsequent step will accumulate over time. Since SpringDTW targets sequence matching by detecting local minima for $d(x_t, y_m)$ it may be prematurely detect incomplete cardiac cycles. Therefore, we show that Boosted-SpringDTW's likelihood scores better distinguish between a sub-optimal subsequence and an optimal cardiac cycle.

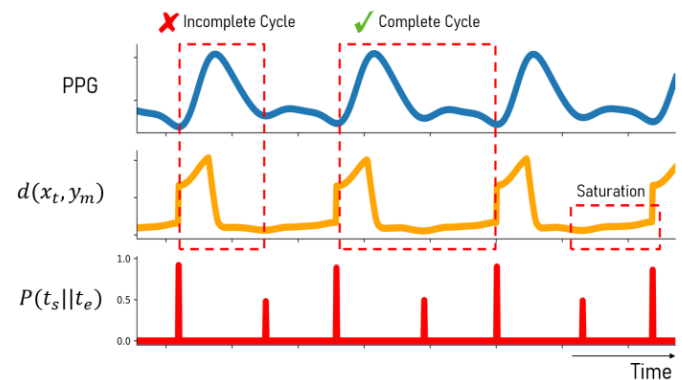

Fig. 7. Challenges associated with $d(x_t, y_m)$ trend saturation are overcome with the $P(t_s \cap t_e)$ endpoint likelihood score based on boosting.

## V. Conclusion

In this work, we proposed Boosted-SpringDTW to perform comprehensive feature extraction for remote health monitoring. We enable the use of DTW for segmentation of quasi-periodic signals and without the need for pre-defined thresholds by combining the strengths of simple, minimal domain-specific heuristics and the generalizable DTW signal analysis method. We overcame the challenges associated with morphological variations in PPG with the notion of a dynamic template. We validated performance by evaluating precision, recall, F1-score, and RMSE performance when attempting to identify SYS, MS, and EP points of a collected PPG signal induced with variation in signal morphology due to inter-subject variability and respiratory behaviors. The proposed framework achieved superior performance for the fiducial point identification task compared to the original SpringDTW implementation and to the standard adaptive thresholding approach. This led to superior IBI estimation by Boosted-SpringDTW frameworks.

## Supplementary Materials

Supplementary Materials are available for download to further elaborate on various concepts referenced in the main manuscript.

# Supplementary Materials

## Boosted-SpringDTW for Comprehensive Feature Extraction of Physiological Signals

Jonathan Martinez*, *Student Member, IEEE*, Kaan Sel, Bobak J. Mortazavi, *Member, IEEE*, and Roozbeh Jafari, *Senior Member, IEEE*

THE Supplementary Materials presented here are divided into two main sections. The first section includes the supplementary text to provide necessary details for the explanations shared in the main manuscript. The second section contains a supplementary figure that supports the analysis.

## VI. SUPPLEMENTARY TEXTS

### A. Photoplethysmography Morphology

PPG sensors consist of a light-emitting diode illuminating the skin and a photodetector that measures the light reflected to it as inversely related to the blood volume present in the sensing area [1]. Its resulting signals are quasi-periodic, corresponding to the contraction and relaxation of the heart that results in two distinct blood pulses that are represented by a systolic peak and a diastolic peak. From these key fiducial points, the most commonly analyzed physiological parameters may be obtained – HR, IBI, and HRV. However, studies have shown value in the remaining points of the cardiac cycle waveform since they may provide more detailed insight into patient health. For example, ratios between the max slope or onset points of the sub-components of a PPG cardiac cycle can yield additional parameters such as vascular aging or arterial stiffness [2]–[4], which may indicate risk for stroke or hypertension. Therefore, the ability to extract a comprehensive set of such features will provide valuable context for healthcare providers to consider when treating patients. However, it is also well known that the morphology of PPG signals is sensitive to several external factors including motion, displacement of the sensor, personal fitness, age, and respiratory behaviors [5]–[7]. This results in fluctuations in the visibility and prominence of each fiducial point in the waveform due to the resulting visibility of the dicrotic notch. For example, the diastolic peak becomes more identifiable with a more prominent dicrotic notch and less identifiable with a less prominent dicrotic notch. Consequently, feature extraction frameworks that do not account for this will yield inaccurate estimations of physiological parameters.

Most previous work develop heuristics based on this understanding to extract these points using the raw PPG signal and its derivatives [8]. In fact, most previous work focused on fiducial point detection depend on this prior knowledge of morphological characteristics [9]–[11]. However, it is often based on adaptive thresholding to capture the general case of waveform morphology – such as, average cycle length, max slope of onset, and minimum height of peaks. Despite this, such approaches enable adaptation only with respect to a pre-defined parameter set that must be carefully established by the researcher, making it difficult to account for all possible sources for variation.

Other previous work has pursued high-quality fiducial point detection by applying transformations to better analyze PPG features in the frequency-domain [12]. Hilbert transformations help with this by enabling more accurate peak detection when a signal experiences high levels of noise, due to its ability to analyze instantaneous frequency and amplitude [13]–[16]. Similarly, wavelet transformations have also been leveraged due to its ability to analyze a signal in both the frequency and time domains [14], [17]–[19]. Other types of filtering and denoising techniques have also been leveraged and have proven effective when faced with a high level of noise imposed by motion. However, such approaches may often still depend on effective adaptive thresholding to identify the targeted fiducial points which is problematic when the noise and signal frequency bands are close to each other since this will require an additional sensing modality (e.g., acceleration) to distinguish between the two. Also, when the signal is contaminated with noise that exists at the same frequency band (e.g., running at 100steps/minute with a HR of 100 beats/minute), adaptive thresholding will again fail to distinguish between the two.

More recently, machine learning and deep learning approaches have been explored for fiducial point detection in not only PPG but also for ECG. Particularly, neural network variations, including long short-term memory (LSTM) and convolutional neural networks (CNN) [20], [21], have proven superior to many other previous work and strong robustness to noise. However, such models rely on a lengthy training process that requires a large dataset of several PPG variations. When a deep learning model is faced with a variation that was not before seen in the training set, it is likely that accuracy of fiducial point identification will suffer. Therefore, the model will have to recognize that a new instance has appeared and must update itself accordingly which is not ideal for the deployable setting where there are limited time and memory resources for retraining the model.

### B. Dynamic Time Warping

DTW serves as the core component of our proposed framework and we adapt it to achieve high-quality segmentation, comprehensive feature extraction, and quality-awareness. The essence of this method is the computation of the distances between all pairs of samples between a target waveform (in our case PPG), $X = (x_1, x_2, x_3, ..., x_n)$, and a template, $Y = (y_1, y_2, y_3, ..., y_m)$, to compose a distance matrix, $D(X, Y)$, of dimensions $m \times n$, where each distance between pairs may be computed as in [22]

$$d(x_t, y_i) = \|x_t - y_i\| + min \begin{cases} d(t, i-1) \\ d(t-1, i) \\ d(t-1, i-1) \end{cases} \quad (1)$$

$$\|x_t - y_i\| = \sqrt{\sum_{j=1}^{|X|,|Y|} (x_{tj} - y_{ij})^2} \quad (2)$$

where $d(x_t, y_i)$ is the distance between pairs in the distance matrix, $\|x_t - y_i\|$ is the Euclidean distance, and $|X|$ and $|Y|$ represent the dimensionality of the waveform and template, $X$ and $Y$.

The DTW distance, $d(x_t, y_i)$, is distinctly characterized by the notion that the difference between sample pairs will be supplemented with the minimum adjacent distance in the matrix. But, the difference between pairs may be computed several different ways to be adapted to the data types being compared and to the problem objective. For raw time-series, the most common distance function to be used is the Euclidean distance which is what we will use in this work. Also, in this work, we will use the first derivative of all signals for comparison as is done in previous work. This is in an effort to improve comparison performance by alleviating the amplitude bias problem and focus comparison on morphology. In addition, we perform mean normalization for all waveforms independently to further enhance performance.

After composition of the distance matrix, the warping path, $w_{X,Y}$, may be extracted which represents the amount of overall alignment between the template and target waveform by directly mapping samples to each other, as shown in Supplementary Fig. II-A. The value of the warping path is computed as:

$$|w_p| = \sum d(x_t, y_m) \quad (3)$$

which is the sum of all pairwise distances along the path. The elements of the path are identified through a traceback starting from the last distance computed in the matrix then moving towards the first distance computed by stepping through indices containing locally minimum distances, as is done in typical dynamic programming algorithms. Previous work has also proposed constraining the search space as a way to make this step more computationally efficient and to prevent unreasonable warping paths. For our framework, we adopt this practice by imposing the Sakoe-Chiba band [23] – shown in Supplementary Fig. II-B – which constrains search to a set amount of deviation from the diagonal of the distance matrix. We conclude this to be effective for our framework since the computation of the warping path will take place after the segmentation task – to be further discussed in a later section. Thus, we expect the detected cardiac cycle and the template to be reasonably in phase.

DTW is a well-established method which has proven its value for comprehensive feature extraction. However, in its original form, DTW is not scale-invariant as it relies on absolute and Euclidean distances between amplitudes of all pairs of samples in each waveform [24]. As morphologies become more complex, DTW will fail due to samples from each waveform that reside in different regions of the morphology may be mapped to each other if they are very close in amplitude. Therefore, there exists many previous works who attempt to improve waveform comparison quality by replacing distance metrics based on raw samples with those that prioritize morphology over scale. The most common early attempt to accomplish this utilized local derivatives and this was later extended to include a number of other shape features – such as piecewise aggregate approximated subsequences, wavelet coefficients, and histogram of oriented gradients – in a framework known as ShapeDTW [25]. The essence of such approaches aim to weight regions of each morphology that are distinctive enough to form a proper alignment, as done with affine and regional DTW or with EventDTW [26], [27]. Although effective, obtaining such shape descriptors is often achieved through windowing or with an iterative approach that will become computationally impractical when being extended to the online analysis of streaming data.

DTW has also been leveraged for subsequence matching, or segmentation, tasks. The naive solution would be to employ an iterative process that continues to all possible subsequence alignments between a template and a segment of a stream until a segment with the minimum overall DTW distance is detected, however, this is computationally inefficient yielding a time complexity of $O(n^3 m)$. Therefore, SpringDTW was proposed in a previous work and is able to accomplish efficient segmentation and fiducial point identification with a time complexity of $O(nm)$ where $m$ is the length of the template signal and $n$ is the length of any given signal to be compared with [22]. This approach depends on a DTW distance threshold to constrain the amount of similarity required for a given segment of a data stream to be detected as an optimal subsequence. However, this will be problematic for data streams whose morphologies evolve over time since this threshold will have to be re-tuned. Also, without careful tuning of this threshold, when analyzing quasi-periodic signals, SpringDTW will detect several incomplete subsequences as false positives. Last, it will be impractical and computationally expensive to incorporate any of the innovations previously proposed to improve DTW comparison quality, such as with shape descriptors.

### C. Dynamic Template - Consensus Sequence

High-quality sequence consensus is essential for generation of a valid dynamic template. Using the detected cardiac cycles of a region of PPG, we leverage DTW barycenter averaging which aims to determine an average morphology of all the waveforms in the set of cardiac cycles by minimizing the DTW distances between them [28], [29]. The length of the resulting consensus sequence may be set as the average cardiac cycle length for the region of PPG, $l_x$. After the newly generated template, $Y_u$, is filtered for smoothing, it will need to be analyzed by $Y^*$ for labeling the locations of its fiducial points. However, if $Y^*$ and $Y_u$ are of different lengths and differ considerably in morphology, the accuracy of the newly labelled fiducial points may suffer. Therefore, we must first resample the current prime template to be of the same length, $l_u$. Then, we compute DTW comparisons between $Y^*$ and the resampled prime template, $Y_r$, to label the fiducial points onto it. Although the lengths of $Y^*$ and $Y_r$ may be different, they have the exact same morphology which should still enable high-accuracy fiducial point identification between these two waveforms. Then, we can compute DTW between $Y_r$ and $Y_u$ to label the new set of fiducial points to be used as ground truth in future analysis. Since the lengths of each of these waveforms is now the same, we can expect a resulting high-quality newly labelled set of fiducial points. It is also important to note that the integrity of the dynamic template is highly dependent on the quality of the prime template. This dependency is considered acceptable since it is the only manually validated template in the ensemble.

### D. Runtime Analysis

The overall runtime for the Boosted-SpringDTW framework using a single template will be $O(nm + \log n + n \log n)$, $O(nm)$ for computing the DTW distance matrix as is done for SpringDTW [22], $O(\log n)$ for peak detection [30] to identify candidate EPs, and $O(n \log n)$ for estimating the average cycle length with FFT [31]. The additional computational cost compared to SpringDTW is a tradeoff for the superior performance in fiducial point identification as shown in Section IV-A. The overall runtime for the Boosted-SpringDTW framework using a dynamic template is $O(knm + \log n + n \log n + eSm^2 + m^2)$, where $k$ is the number of iterations required for analysis by each template in the ensemble, $O(eSm^2)$ for updating the dynamic template with DTW barycenter averaging [29], where $e$ is the number of iterations set to optimize the average waveform, $S$ is the number of cardiac cycles in the 1-minute window, $m$ is the defined length of the newly generated template, and $O(m^2)$ for automatically labelling the new template with the standard DTW algorithm – since both waveforms will be the same length. While this does yield a significant increase in runtime compared to the original implementation of SpringDTW, the results discussed in Section IV-A and IV-B show that the tradeoff is worth the large increase in fiducial point identification and physiological parameter estimation performance.

## VII. SUPPLEMENTARY FIGURES

### A. Dynamic Time Warping

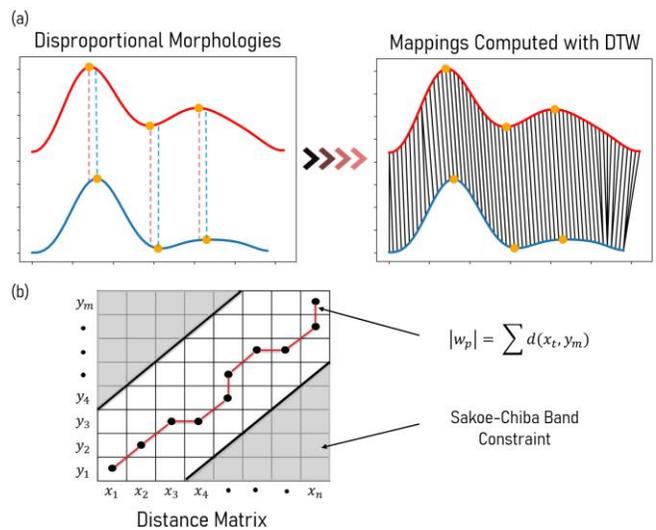

Supplementary Fig. A. Illustration of signal mappings, DTW distance matrix, and extraction of the warping path.